\documentclass{iopart}
\usepackage{epsfig}
\begin{document}
\title
{
Nonequilibrium dynamics of a growing interface
}
\author{Hans C Fogedby}
\address{
\footnote[1]{Permanent address}Institute of Physics and Astronomy,
University of Aarhus, DK-8000, Aarhus C, Denmark
\\
and
NORDITA, Blegdamsvej 17, DK-2100, Copenhagen {\O}, Denmark
}
\begin{abstract}
A growing interface subject to noise is described by the
Kardar-Parisi-Zhang equation or, equivalently, the noisy 
Burgers equation. In one dimension this equation is analyzed 
by means of a weak noise canonical phase space approach
applied to the associated Fokker-Planck equation. The growth 
morphology is characterized by a gas of nonlinear soliton modes 
with superimposed linear diffusive modes. We also discuss
the ensuing scaling properties.
\end{abstract}
\section{Introduction}
Macroscopic phenomena far from equilibrium abound and include
phenomena such as turbulence in fluids, interface and growth problems,
chemical reactions, processes in glasses and amorphous systems, 
biological processes, and even aspects of economical
and sociological structures.

In recent years much of the focus of modern statistical physics
and soft condensed matter has shifted towards such systems. Drawing on 
the case of static and dynamic critical phenomena in and close to
equilibrium, where scaling, critical exponents, and the concept of
universality have so successfully served to organize our understanding
and to provide a variety of calculational tools, a similar approach
has been advanced towards the much larger class of nonequilibrium
phenomena with the purpose of elucidating scaling properties and more
generally the morphology or pattern formation in a driven state.

In this context the noisy Burgers equation or the equivalent
Kardar-Parisi-Zhang (KPZ) equation, describing the nonequilibrium growth
of a noise-driven interface, provide simple continuum models
of an open driven nonlinear system exhibiting scaling and pattern
formation.

\begin{figure}
\begin{center}
\epsfxsize=5cm
\epsfbox{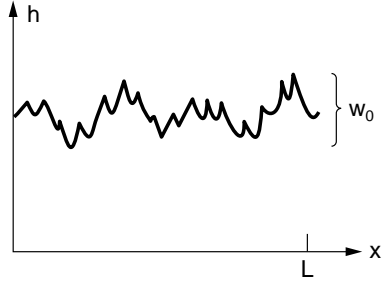}
\end{center}
\caption{
We depict the growth morphology of a growing interface in a system
of size $L$. The saturation width in the noise-driven stationary
state is denoted by $w_0$.
}
\label{fig1}
\end{figure}

In one dimension the KPZ equation has the form (in a co-moving frame)
\cite{Kardar86,Medina89}
\begin{eqnarray}
\frac{\partial h}{\partial t} = 
\nu\nabla^2 h + \frac{\lambda}{2}(\nabla h)^2 + \eta ~.
\label{kpz}
\end{eqnarray}
Here $h$ is the height field, $\nu$ a damping or 
viscosity characterizing the linear diffusive term, 
$\lambda$ a coupling strength for the nonlinear mode coupling or
growth term, and $\eta$ a Gaussian white noise, driving the system into a 
stationary state. The noise is correlated according to
\begin{eqnarray}
\langle\eta(xt)\eta(00)\rangle = \Delta\delta(x)\delta(t)~,
\label{noise}
\end{eqnarray}
and characterized by the noise strength $\Delta$.
In figure \ref{fig1} we have depicted a realization of a growing interface.

\begin{figure}
\begin{center}
\epsfxsize=5cm
\epsfbox{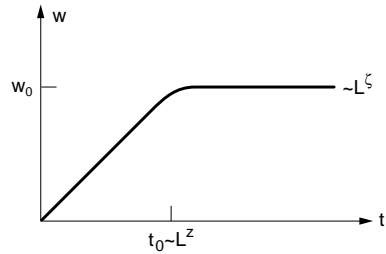}
\end{center}
\caption{
We depict the interface width $w(t)$ as a function of
$t$. In the transient regime for $t\ll t_0\sim L^z$
$w$ grows according to $t^{\zeta/z}$. In the stationary regime
attained for
$t\gg t_0$ the width $w$ saturates to the value $w_0\sim L^{\zeta}$.
}
\label{fig2}
\end{figure}

Assuming an initially flat interface the width $w(t,L)$ grows in time
approaching the saturation width $w_0$ in the stationary regime. The
dynamical scaling hypothesis
\cite{Halpin95,Barabasi95} then asserts that
\begin{eqnarray}
w(t,L)=L^\zeta F_{\rm w}(t/L^z)~,
\label{wscal}
\end{eqnarray}
where $\zeta$ is the roughness exponent characterizing the morphology
of the interface and $z$ the dynamic exponent describing the dynamical
correlations. The scaling function $F_{\rm w}(u)$ together with
$\zeta$ and $z$ define the scaling universality class and has the
limits $F_{\rm w}(u)\rightarrow\rm {const.}$ for 
$u\rightarrow\infty$ and $F_{\rm{w}}(u)\rightarrow u^{\zeta/z}$ for
$u\rightarrow 0$.
In figure \ref{fig2} we have shown the time- and size dependent 
width $w(t,L)$
Similarly, applying the dynamical scaling hypothesis to the height
correlations in the stationary regime we have
\begin{eqnarray}
\langle(h(xt)-\langle h(xt)\rangle)(h(00)-\langle h(00)\rangle)\rangle
=
x^{2\zeta}F_{\rm{h}}(t/x^z) ~,
\label{hscal}
\end{eqnarray}
where the scaling function $F_{\rm{h}}$ obeys
$F_{\rm{h}}(u)\rightarrow\rm{const.}$ for
$u\rightarrow\infty$ and $F_{\rm{h}}(u)\rightarrow u^{2\zeta/z}$ for
$u\rightarrow 0$.

\begin{figure}
\begin{center}
\epsfxsize=5cm
\epsfbox{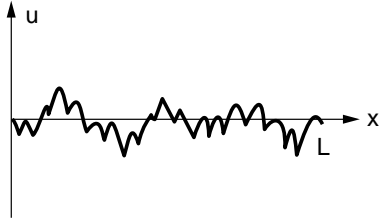}
\end{center}
\caption{
We depict the morphology of a growing interface
in terms of the local slope. The slope field  fluctuates about zero.
}
\label{fig3}
\end{figure}

In what follows it turns out that the local slope field 
$u=\nabla h$ is the appropriate variable in which to discuss the
growth morphology of an interface. The height field
$h=\int dx~u$ is then an integrated variable sampling the slope 
fluctuations. In figure \ref{fig3} we have depicted a realization
of the slope field
The dynamical scaling hypothesis for the slope field then reads
\begin{eqnarray}
\langle u(xt)u(00)\rangle
=
x^{2\zeta-2}F_{\rm{u}}(t/x^z) ~,
\label{uscal}
\end{eqnarray}
with limits
$F_{\rm{u}}(u)\rightarrow\rm{const.}$ for
$u\rightarrow\infty$ and $F_{\rm{u}}(u)\rightarrow u^{(2\zeta-2)/z}$ for
$u\rightarrow 0$.

The stochastic dynamics of the slope field is then according to 
Eq. (\ref{kpz}) governed by the noisy Burgers equation
\cite{Forster76,Forster77}
\begin{eqnarray}
\frac{\partial u}{\partial t} = \nu\nabla^2u + \lambda u\nabla u +\nabla\eta ~.
\label{bur}
\end{eqnarray}
This equation in the noiseless case for $\eta = 0$ was originally proposed
by Burgers \cite{Burgers74} in order to model turbulence in fluids;
note the similarity with the Navier-Stokes equation for $\lambda = -1$.

The substantial conceptual problems encountered in nonequilibrium
physics
are in many ways embodied in the KPZ-Burgers equations (\ref{kpz}) and
(\ref{bur})
describing the self-affine growth of an interface subject to
annealed noise arising from fluctuations in the drive or in
the environment, and these equations
in one and
higher dimensions have  been the subject of intense scrutiny in recent
years
owing to their paradigmatic significance within a field theory 
of nonequilibrium
systems
\cite{Halpin95,Barabasi95,Krug97,Frey94,Frey96,Frey99,Janssen99,Laessig98a,Laessig00,Colaiori01a,Colaiori01b}.

Interestingly, the Burgers-KPZ equations are also encountered in a variety
of other problems such as randomly stirred fluids,
dissipative transport in a driven lattice gas,
the propagation of flame fronts,
the sine-Gordon equation, and magnetic
flux lines in superconductors. Furthermore, by means of the 
Cole-Hopf
transformation the Burgers-KPZ equations are also
related to
the problem of a directed polymer or
a quantum particle in a random
medium  and thus to the theory of spin glasses.

In a series of papers the one dimensional case
defined by (\ref{bur}) 
has been analyzed in an attempt to uncover
the
physical mechanisms underlying the pattern formation and scaling
behavior.
Emphasizing that the noise strength $\Delta$ constitutes the  relevant 
nonperturbative parameter driving the system into a 
statistically stationary state,
the method was initially based on a weak noise saddle point
approximation to the Martin-Siggia-Rose functional formulation
of the
noisy Burgers equation \cite{Martin73,Baussch76,Fogedby98b,Fogedby98c}.
This work was a continuation of earlier work based on the mapping
of a solid-on-solid model onto a continuum spin model
\cite{Fogedby95}.
More recently the functional approach has been
superseded by a {\em canonical phase space method} 
deriving from the 
canonical structure of the Fokker-Planck 
equation associated with the
Burgers equation
\cite{Freidlin84,Graham84,Fogedby99a,Fogedby99b}.
In the present context we attempt to give a brief account of
this approach with emphasis on the physical interpretation.
\section{The noisy Burgers equation}
The noisy Burgers equation (\ref{bur}) has the form of a conserved
nonlinear Langevin equation, $\partial u/\partial t = -\nabla j$,
with fluctuating current $j = -\nu\nabla u - (\lambda/2) u^2 + \eta$.
The hydrodynamical origin of the Burgers equation, 
as reflected by the presence of the mode coupling or convective
term $\lambda u\nabla u$, 
implies that the Burgers equation is invariant under the Galilean 
transformation
\begin{eqnarray}
x\rightarrow x-\lambda u_0t~~{\rm{and}}~~ u\rightarrow u+u_0 ~,
\label{gal}
\end{eqnarray}
involving a shift of the slope field.
Since the nonlinear
coupling strength $\lambda$ enters as a structural constant
in the symmetry group it is invariant under scaling and a simple
Kadanoff-type  block renomalization group argument in both
space and time implies
the dynamical scaling law
\begin{eqnarray}
\zeta + z = 2 ~,
\label{scal}
\end{eqnarray}
providing a relationship between the roughness
exponent and the dynamic exponent.

In the linear case for $\lambda=0$ the Burgers equation
takes the Edwards-Wilkinson (EW) form \cite{Edwards82}
\begin{eqnarray}
\partial u/\partial t = \nu\nabla^2u + \nabla\eta ~ ,
\label{ew}
\end{eqnarray}
i.e., a linear diffusion equation driven by conserved noise.
This equation is easily decomposed and analyzed in terms of 
wavenumber modes
$u_k = \int dx~ u(x)\exp{(-ikx)}$, $u_k^{\ast}=u_{-k}$. In the noiseless
case for $\eta=0$ the field $u_k$, governed by an overdamped oscillator
equation, decays according to $u_k\propto\exp(-\nu k^2t)$;
$\tau_k=1/\nu k^2$  setting the spectrum of relaxation times. 
For $\eta\neq 0$
the slope field $u_k$ is driven into a noisy stationary state.
Defining $u_{k\omega} =\int dt~u_k\exp(i\omega t)$ the stationary
correlations are given by
\begin{eqnarray}
\langle u_{k\omega}u_{-k-\omega}\rangle = 
\frac{\Delta k^2}
{\omega^2 + (\nu k^2)^2} ~.
\label{cor0}
\end{eqnarray}
We note that (\ref{cor0}) in addition to the pole $\omega = i\nu k^2$,
corresponding to the decaying mode, also has a pole at 
$\omega = -i\nu k^2$, characterizing a growing mode. The noise in
driving the equation thus excites a growing diffusive mode and we have
generally for a particular realization
\begin{eqnarray}
u(xt)=(Ae^{-\nu k^2t}+Be^{+\nu k^2t})~e^{ikx} ~.
\label{mode}
\end{eqnarray}
In the stationary state the correlation are time reversal invariant,
requiring both growing and decaying modes; this  also follows from
the Fokker-Planck analysis in section 3.
From (\ref{cor0}) we also obtain the static correlations
$\langle u(x)u(0)\rangle=\Delta/2\nu\delta(x)$,
showing that $u(x)$ is spatially uncorrelated. Moreover, the stationary
distribution has the form
\begin{eqnarray}
P_{\rm{st}}(u)\propto\exp{\left[-\frac{\nu}{\Delta}\int dx~
u(x)^2\right]}~.
\label{sdis}
\end{eqnarray}
Comparing (\ref{cor0}) with the scaling form (\ref{uscal})
we also infer the scaling exponents $\zeta
=1/2$ and $z=2$, 
characteristic of diffusion and defining the EW universality class.
Also, noting that the diffusive term in  (\ref{ew}) can be
derived from a free energy $F = (1/2)\int dx u^2$, it follows that
the EW equation  describes the fluctuations in an equilibrium
system with temperature $\Delta/2\nu$, i.e.,
$P_{\mbox{st}} = \exp{[-(2\nu/\Delta)F]}$.

Summarizing, in the linear case the elementary excitations are decaying
and growing nonpropagating diffusive modes. Since from (\ref{kpz})
$\partial\langle h\rangle/\partial t = \nu\nabla^2\langle h\rangle$
the average height $\langle h\rangle$ decays to zero 
(with respect to the co-moving
frame) and the diffusive 
modes do not represent nonequilibrium growth but  
in fact characterizes equilibrium fluctuations at the 
temperature $\Delta/2\nu$.
Finally, the exponents $\zeta=1/2$ and $z=2$ for the EW
universality class  reflect the
uncorrelated slope field, or the random walk of the integrated slope,
i.e., the height $h$, and the
diffusional character of the dynamics, 
respectively.

\begin{figure}
\begin{center}
\epsfxsize=8cm
\epsfbox{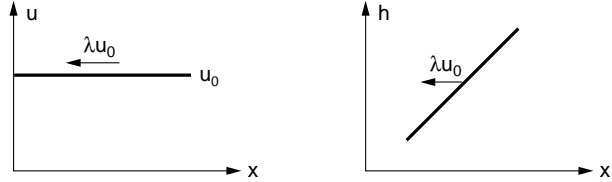}
\end{center}
\caption{
We show the slope and height fields for a constant
shift $u_0$. The sloped height profile then moves
with velocity $\lambda u_0$.
}
\label{fig4}
\end{figure}

\begin{figure}
\begin{center}
\epsfxsize=8cm
\epsfbox{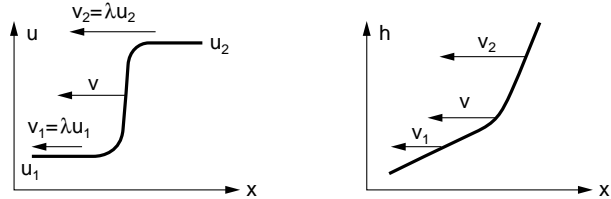}
\end{center}
\caption{
We show a step profile in the slope field with amplitudes
$u_1$ and $u_2$. The step moves with mean velocity
$v=(v_1+v_2)/2$, where $v_1=\lambda u_1$ and $v_2=\lambda u_2$
are the velocities of the plateaus. The mode correspond to
the slope dependent velocity of the height field subject to growth.
}
\label{fig5}
\end{figure}

In the nonlinear case for $\lambda\neq 0$ the Galilean invariance
(\ref{gal}) becomes operational and we can draw some simple conclusions
concerning a growing interface.
First, imposing a constant shift $u_0$ (\ref{gal}) implies that
the tilted height field propagates with velocity $\lambda u_0$ as
shown in figure \ref{fig4}. 
Next, considering a step in the slope field
with amplitudes $u_1$ and $u_2$ propagating with velocities
$v_1=\lambda u_1$ and $v_2=\lambda u_2$, it follows that the step itself 
propagates with mean velocity $v=(v_1+v_2)/2$. For the height field
this corresponds to the part with the largest slope moving with
the largest velocity in accordance with the slope dependent growth velocity
in the KPZ equation (\ref{kpz}), i.e.,  
$\partial\langle h\rangle/\partial t=
(\lambda/2)\langle(\nabla h)^2\rangle>0$.
That kind of localized modes 
giving rise to growth are indeed supported by the Burgers equation
which possesses a spectrum of soliton modes.
The configuration is depicted in figure \ref{fig5}.

In the noiseless case for $\eta=0$ the Burgers equation support a right
hand single soliton solution, in the static case of the kink-like form
\begin{eqnarray}
u(x)=u_0\tanh\frac{\lambda u_0}{2\nu}(x-x_0)~,
\label{sol1}
\end{eqnarray}
localized at $x_0$, with amplitude $u_0$, and width $2\nu/\lambda u_0$.
Boosting the soliton to a finite velocity this mode corresponds to the
configuration depicted in figure \ref{fig5}. Denoting the right and left
hand boundaries by $u_+$ and $u_-$, respectively, we also infer
the soliton condition
\begin{eqnarray}
u_+ + u_-=-2v/\lambda ~.
\label{solcon}
\end{eqnarray}
The relaxing growth morphology of the noiseless Burgers 
equation corresponding to the deterministic transient
growth of the KPZ equation can be described by a gas of propagating
right hand solitons, connected by constant-slope ramp solutions, and
with a spectrum of superimposed decaying linear modes. In the height
field this morphology corresponds to downward cusps connected by
parabolic segments with superimposed linear modes 
\cite{Kardar86,Medina89,Fogedby98a,Fogedby01a}.

In the noisy case the interface is driven into a stationary state,
and anticipating the analysis in section 3, it turns out that the noise excites
a left hand soliton of the shape
\begin{eqnarray}
u(x)=-u_0\tanh\frac{\lambda u_0}{2\nu}(x-x_0)~.
\label{sol2}
\end{eqnarray}
This mode is a solution of the growing noiseless Burgers equation for
$\nu\rightarrow -\nu$. This doubling of modes is equivalent to
the linear case where the mode has the form (\ref{mode}). The stationary
growth morphology can thus be described by a gas of right hand and
left hand solitons matched by the soliton condition (\ref{solcon})
with superposed linear modes.

The Fokker-Planck equation for the Burgers equation has the form
\begin{eqnarray}
\Delta\frac{\partial P}{\partial t} = HP~,
\label{fp}
\end{eqnarray}
where $P$ is the probability distribution and $H$ a Hamiltonian
(Liouvillean) driving the equation of the form
\begin{eqnarray}
H=&&-\Delta\int dx\frac{\delta}{\delta u}(\nu\nabla^2u+
\lambda u\nabla u)
\nonumber
\\
&&+\frac{1}{2}\Delta^2\int
dxdx'~\nabla\nabla'\delta(x-x')
\frac{\delta^2}{\delta u\delta u}~.
\label{ham}
\end{eqnarray}
Whereas (\ref{fp} will be discussed in more detail in section 3
it follows easily  that the stationary
solution of (\ref{fp}) has the form in (\ref{sdis}) \cite{Huse85}, 
independent
of $\lambda$, and we infer as in the linear case $\zeta = 1/2$.
The scaling law (\ref{scal}) then implies the dynamic exponent
$z=3/2$. 

Summarizing, in the Burgers case the elementary excitations are 
propagating
right hand and left hand solitons with superimposed linear modes.
The soliton propagation corresponds to nonequilibrium growth.
Finally, the scaling exponents $\zeta=1/2$ and $z=3/2$ defining
the Burgers/KPZ universality class correspond
to the uncorrelated slope field or random walk of $h$ and 
soliton propagation,
respectively.
\section{The weak noise approach}
Apart from numerical modeling and analysis of other models falling
in the same universality class, the standard analytical approaches
to the noisy Burgers equation are i) the dynamical renormalization
group method (DRG) and ii) the mode coupling approach (MC). The DRG
method accesses the scaling regime and provide an epsilon expansion
about $d=2$. Above $d=2$ the system exhibits a kinetic phase
transition from a smooth phase with EW exponents to a rough phase
controlled by a strong coupling fixed point with largely unknown
scaling exponents. In $d=1$ the scaling is controlled by a strong 
coupling fixed point and the DRG yields (fortuitously) the known
scaling exponents. Unlike the success of the DRG in dynamical critical
phenomena, the results obtained for the noisy Burgers equation
are limited despite a substantial theoretical effort.
The MC method by neglecting (unrenormalized)
vertex corrections provides closed equations 
yielding scaling functions.
However, the somewhat ad hoc nature of the MC approach makes it difficult
to make contact with more systematic approaches.  

The functional or the equivalent phase space approach valid in the
weak noise limit, $\Delta\rightarrow 0$, replaces the stochastic
Langevin-type Burgers equation (\ref{bur}) by coupled deterministic
diffusion-advection type mean field equations,
\begin{eqnarray}
\frac{\partial u}{\partial t}&&=
\nu\nabla^2 u -\nabla^2p+\lambda u\nabla u ~,
\label{mfe1}
\\
\frac{\partial p}{\partial t}&&=
-\nu\nabla^2 p +\lambda u\nabla p ~,
\label{mfe2}
\end{eqnarray}
for the slope $u(x,t)$ and a canonically conjugate noise field
$p(x,t)$, replacing the stochastic noise $\eta(x,t)$. The field equations
bear the same relation to the Fokker-Planck equation 
(\ref{fp}) as the classical
equations of motion bear to the Schr\"{o}dinger equation in the 
semi-classical WKB approximation.

To justify the weak noise limit we recall the
analogy with the WKB approximation in quantum mechanics which, owing to
its nonperturbative character, captures features like 
bound states and tunneling amplitudes, which are generally inaccessible
to perturbation theory. Therefore, we anticipate that the present
weak noise approach to the Burgers equation also accounts 
correctly, at least in a qualitative sense, for the stochastic properties
even at larger noise strength. However, there may be an upper
threshold value beyond which the
system may enter a new stochastic or kinetic
phase. In the one dimensional case discussed here
the scaling behavior is controlled by a single
strong coupling fixed point which can be accessed
by the present weak noise approach. In two and higher
dimension a dynamic renormalization group analysis
predicts a kinetic phase transition at a critical noise
strength (or coupling strength) and the weak noise
approach presumably fails. 

The equations (\ref{mfe1}) and (\ref{mfe2}) derive from
a principle of least action characterized 
by an action $S(u'\rightarrow u'',t)$ for an 
orbit $u'(x)\rightarrow u''(x)$ traversed in time $t$
\begin{eqnarray}
S(u'\rightarrow u'',t) = \int_{0,u'}^{t,u''}dt\,dx
\left(p\frac{\partial u}{\partial t} - {\cal H}\right) ~,
\label{act}
\end{eqnarray}
with Hamiltonian density
\begin{eqnarray}
{\cal H} = p\left(\nu\nabla^2u + \lambda u\nabla u - 
\frac{1}{2}\nabla^2 p\right) ~.
\label{hamden}
\end{eqnarray}
The action is of central importance in the present approach
and serves as a  weight function for the 
noise-driven
nonequilibrium configurations in much the same
manner as the energy $E$ in the Boltzmann factor $\exp(-\beta E)$ for
equilibrium systems, where $\beta$ is the inverse temperature. 
The dynamical action in fact
replaces the energy in the context of the dynamics of stochastic 
nonequilibrium systems governed by a generic Langevin equation
driven by Gaussian white noise.
The action provides a methodological approach and 
yields access to the
time dependent and stationary probability distributions,
\begin{eqnarray}
&&P(u'\rightarrow u'',t)\propto
\exp\left[-\frac{S(u'\rightarrow u'',t)}{\Delta}\right] ~,
\label{dis} 
\\
&&P_{\rm{st}}(u'') = \lim_{t\rightarrow\infty}P(u'\rightarrow u'',t) ~,
\label{statdis}
\end{eqnarray}
and associated moments, e.g., the stationary slope correlations
\begin{eqnarray}
\langle u(xt)u(00)\rangle = 
\int\prod du~ u''(x)u'(0)P(u'\rightarrow u'',t)P_{\rm{st}}(u') ~.
\label{cor}
\end{eqnarray}

The contact with the Fokker-Planck equation (\ref{fp}) in the
weak noise limit is established by noting that (\ref{dis})
inserted in (\ref{fp}) to leading order yields the
Hamilton-Jacobi equation $\partial S/\partial t+H(p,u)=0$ 
with canonical momentum
$p=\delta S/\delta u$, where the Hamiltonian $H=\int dx~{\cal H}$.

The canonical formulation associates the conserved energy $E$
(following from time translation invariance), the conserved
momentum $\Pi$ (from space translation invariance), and the 
conserved area $M$ (from the Burgers equation with conserved
noise):
\begin{eqnarray}
&&E=\int dx\,{\cal H}~,
\label{ener}
\\
&&\Pi = \int dx\, u\nabla p~,
\label{mom}
\\
&&M = \int dx\, u~.
\label{area}
\end{eqnarray}

\begin{figure}
\begin{center}
\epsfxsize=8cm
\epsfbox{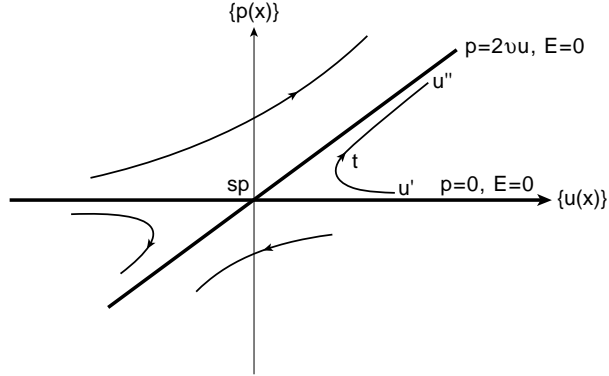}
\end{center}
\caption{
Generic behavior of the orbits in $up$ phase space.
Heavy lines indicate the zero energy manifold. The stationary
saddle point (sp) is at the origin. The finite time orbit from
$u'$ to $u''$ is attracted to the saddle point for
$t\rightarrow\infty$.
}
\label{fig6}
\end{figure}

The field equations (\ref{mfe1}) and (\ref{mfe2}) determine orbits
in a canonical $up$ phase space where the dynamical issue in
determining $S$ and thus $P$ is to find an orbit from $u'$ to
$u''$ in time $t$, $p$ being a slaved variable. Note that unlike
dynamical system theory we are not considering the asymptotic
properties of a given orbit. In general 
the orbits in phase space lie on the manifolds
determined by the constants of motion $E$, $\Pi$ and $M$. Here the
zero energy manifold $E=0$ plays a special role in defining the
stationary state. For vanishing or periodic boundary conditions
for the slope field the zero energy manifold is  
composed of the
transient submanifolds $p=0$ and the stationary submanifold 
$p=2\nu u$.
The zero energy
orbits on the $p=0$ manifold correspond to solutions of the
damped noiseless Burgers equation; the orbits on the $p=2\nu u$
are solutions to the undamped noiseless Burgers equation with negative
damping, i.e., $\nu$ replaced by $-\nu$. In the solvable linear case of the
noise driven diffusion equation for $\lambda=0$, i.e., the Edwards-Wilkinson
equation \cite{Edwards82}, a finite 
energy orbit from $u'\rightarrow u''$ in time $t$
migrates to the zero energy manifold in the limit $t\rightarrow\infty$,
yielding according to (\ref{act}) and (\ref{statdis}) the stationary
distribution $P_{\rm{st}}\propto\exp(-(\nu/\Delta)\int dx\, u^2)$.
This distribution also holds in the Burgers case and is a generic result
independent of $\lambda$ \cite{Huse85}. Finally, in the long time 
limit an orbit from $u'\rightarrow u''$ is attracted to the 
hyperbolic saddle point
at the origin in phase space implying ergodic behavior in the stationary
state. In figure \ref{fig6} we have schematically depicted 
possible orbits in phase
space.

In the linear case the field equations (\ref{mfe1}) and (\ref{mfe2})
couple $p$ parametrically to $u$, i.e. $p$ is slaved. In wavenumber
space $p_k$ is growing, whereas $u_k$ driven by $p_k$ is a linear
superposition of damped and growing diffusive modes, supporting
the expression (\ref{mode}).
In the nonlinear case (\ref{mfe1}) and (\ref{mfe2}) admit nonlinear
soliton or smoothed shock wave solutions which are, in the static case, of
the kink-like form in (\ref{sol1}) and (\ref{sol2}).
Propagating solitons are subsequently generated by the Galilean boost 
(\ref{gal}) and we recover the soliton condition (\ref{solcon}).

\begin{figure}
\begin{center}
\epsfxsize=8cm
\epsfbox{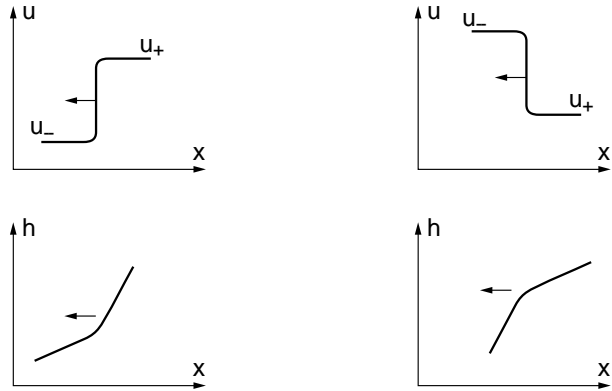}
\end{center}
\caption{
Slope field $u$ and height profile $h$ for
the right hand and left hand moving kink solitons 
in the description of a growing interface.
}
\label{fig7}
\end{figure}

The right hand soliton
moves on the
noiseless manifold $p=0$ and is also a solution of the damped
(stable) noiseless
Burgers equation for $\eta =0$. The noise-induced left hand
soliton 
is associated with the noisy manifold $p=2\nu u$, 
and is a solution of the undamped
(unstable) noiseless Burgers equation with $\nu$ replaced by
$-\nu$. 
In addition the field equations also 
admit linear mode solutions superimposed 
as ripple modes on the solitons.
The ripple modes are superpositions of both decaying and growing
components reflecting the noiseless and noisy manifolds $p=0$ and
$p=2\nu u$, respectively.
The soliton mode induces a propagating component with
velocity $\lambda u$ in such a way that the right hand soliton acts
like a sink and the left hand soliton as a source of linear modes
\cite{Fogedby01a}.
This mechanism will be discussed heuristically in section 4.
In the Edwards-Wilkinson limit for $\lambda\rightarrow 0$ the ripple
modes become the usual diffusive modes (growing and decaying) of the
driven stationary diffusion equation.
In figure \ref{fig7} we have shown the right hand and left hand
solitons.

The heuristic physical picture that emerges from our analysis now
supported by a weak noise analysis of the Fokker-Planck equation is that of a 
many body formulation
of the pattern formation of a growing interface in terms of a dilute
gas of propagating solitons matched according to the soliton condition
(\ref{solcon}) with superimposed linear ripple modes. 
\section{A growing interface}
\begin{figure}
\begin{center}
\epsfxsize=6cm
\epsfbox{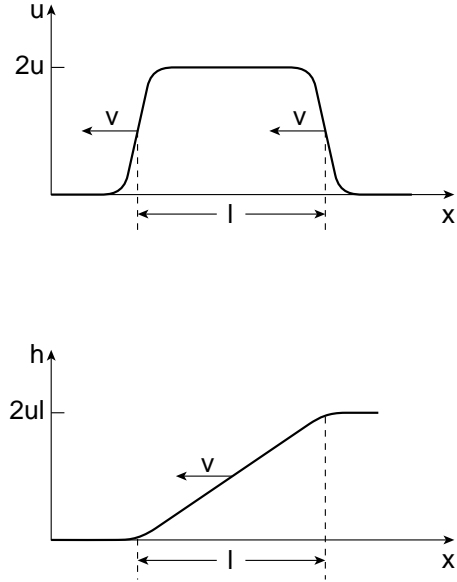}
\end{center}
\caption{
Slope field 
and the resulting height profile
for a soliton pair configuration.
}
\label{fig8}
\end{figure}

As discussed above a growing interface can be envisaged as a many body
system in the Landau quasi particle sense composed of matched right hand
and left hand solitons with superimposed linear modes. Focusing on the 
solitons the right and left hand kinks are the fundamental growth modes
corresponding to cusps in the height field; they are the quarks in 
the present formulation. They do, however, not satisfy periodic 
or vanishing boundary
conditions in the slope field $u$; the nonvanishing boundary
values $u_+$ and $u_-$ in fact correspond to a deterministic current 
dissipated or generated at the soliton centers yielding 
permanent profile solutions. The simplest mode satisfying periodic 
boundary conditions is the  two-soliton or pair soliton configuration
obtained by matching a right hand and a left hand soliton boosted to the
velocity $v=-\lambda u$. The two-soliton mode has amplitude $2u$ and
size $\ell$. 
By inspection it is seen that the pair mode is an
approximate solution to the field equations (\ref{mfe1}) and
(\ref{mfe2}). The correction terms are of the type $u\nabla u$ and
$u\nabla p$ 
and thus correspond to local perturbations from a region of size
$\nu/\lambda|u|$ which is small in the low viscosity limit
$\nu\rightarrow 0$. We assume that the correction can be treated within
a linear stability analysis and thus gives rise to a linear mode
propagating between the 
right hand and left hand solitons. This property is borne out by
a recent numerical analysis of the field equations \cite{Fogedby01b}.
As also shown in the numerical analysis 
the pair mode forms a long-lived excitation or quasi-particle 
in the many body description of a growing interface. Subject to periodic
boundary conditions this mode corresponds to a simple growth situation.
The propagation of the pair mode corresponds to the propagation of a
step in the height field $h$. At each revolution of the pair mode the
interface grows by a uniform layer of thickness $2u\ell$. In 
figure \ref{fig8} we have depicted the pair mode
and the associated height profile $h$.

\begin{figure}
\begin{center}
\epsfxsize=9cm
\epsfbox{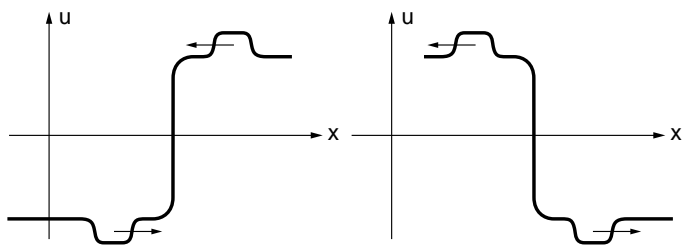}
\end{center}
\caption{
Source and drain mechanism for the right hand and left
hand solitons. The perturbation attracted and repelled by the
soliton centers are modeled by pair solitons.
}
\label{fig9}
\end{figure}

The soliton picture also allows us easily to understand in what sense
the right hand soliton acts like a drain and the left hand soliton
as a source with respect to perturbations. 
Considering two pair solitons
superimposed on the right and left horizontal parts of the static solitons
(\ref{sol1}) and (\ref{sol2}) it follows from (\ref{solcon}) 
that for a right hand
soliton perturbations move toward the soliton center
and for a left hand soliton perturbations move
away from the soliton center. This mechanism also follows from the
linear analysis of ripple modes \cite{Fogedby01a}. 
The mechanism is depicted
in figure \ref{fig9}.

\begin{figure}
\begin{center}
\epsfxsize=6cm
\epsfbox{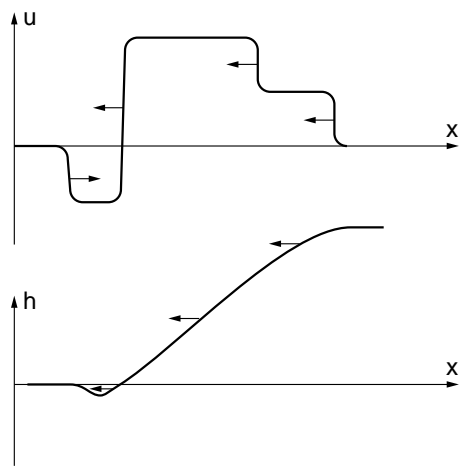}
\end{center}
\caption{
Four soliton representation of the slope field
and the height profile.
}
\label{fig10}
\end{figure}

Generally a growing interface, ignoring the superimposed linear
ripple modes, can at a given time instant be represented by a
gas of matched left hand and right hand solitons as depicted in
figure \ref{fig10} in the four soliton case. A gas of pair solitons
thus constitute
a particular growth mode where the height profile between moving
steps has horizontal segments.
\section{Dynamics, statistics, and scaling}
There two levels of description:
the stochastic Langevin level and the deterministic Fokker-Planck
or equations of motion level. On the Fokker-Planck level yielding
the canonical field equations (\ref{mfe1}) and (\ref{mfe2}) the
growth of the interface is interpreted in terms of a gas of
propagating solitons (and diffusive modes). The stochastic
description on the Langevin level is then established in the
weak noise limit $\Delta\rightarrow 0$ by computing the action $S$
associated with a particular dynamical mode and subsequently deduce
the probability distribution according to (\ref{dis}), i.e.,
$P\propto\exp(-S/\Delta)$. This procedure is completely equivalent
to the WKB limit of quantum mechanics. Here the wavefunction
$\Psi$ and thus the probabilistic interpretation is given by
$\Psi\propto\exp(iS/\hbar)$, where $S$ is the action associated
with the classical motion. Note that unlike quantum
mechanics
there is no phase interference in the stochastic nonequilibrium
case.
\subsection{Dynamics}
\begin{figure}
\begin{center}
\epsfxsize=4cm
\epsfbox{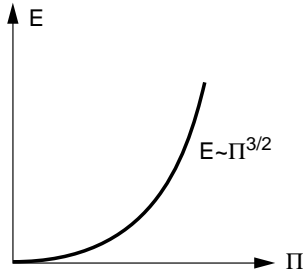}
\end{center}
\caption{
Dispersion law for the noise-induced left hand soliton.
}
\label{fig11}
\end{figure}

The canonical phase space approach discussed in section 3
associates a formal dynamics with the soliton-diffusive mode
gas representation of a growing interface. According to
(\ref{ener}), (\ref{mom}), and (\ref{area}) the total energy,
momentum and area are conserved in the course of the dynamical
evolution of an interface on the deterministic Fokker-Planck
level as determined by the field equations (\ref{mfe1}) and (\ref{mfe2}).
Since the symplectic structure via the Fokker-Planck equation
is associated with the noisy case only the noise-induced
left hand soliton and, similarly, the growing linear mode, 
carries dynamical attributes. By insertion
the static soliton (\ref{sol2}) thus has the energy 
$E=-(16/3)\lambda\nu u_0^3$ and momentum $\Pi =0$. Correspondingly,
the moving pair soliton excitation depicted in figure \ref{fig8} is endowed
with the same energy but has momentum $\Pi=4\nu u|u|$
pointing in the same direction as the velocity, it moreover conserves
the area under propagation. Eliminating the amplitude dependence
the dynamical characteristics of the left hand soliton is conveniently
given by the soliton dispersion law
\begin{eqnarray}
|E| = \frac{2}{3}\frac{\lambda}{\nu^{1/2}} \Pi^{3/2}~.
\label{dislaw}
\end{eqnarray}
We note the fractional power arising from
the nonlinear character of the soliton; the
dispersion law is  shown in figure \ref{fig11}.
\subsection{Statistics}
We consider here as a simple application the statistics of the
long-lived pair soliton shown in figure \ref{fig8}. It has size 
$\ell$, amplitude $2u$, and propagates with velocity
$v=-\lambda u$. During a revolution in a system of size $L$ with
periodic boundary conditions the height field increases with a
layer of thickness $2u\ell$. Since the system is traversed in time
$t=L/v$ the integrated growth velocity is given by
$2\lambda u^2\ell/L$ which for a single pair of fixed size vanishes in
the
thermodynamic limit. On the other hand,the local growth 
velocity $dh/dt$ is given by
$2\lambda u^2=(\lambda/2)(\nabla h)^2$ which is consistent with the
averaged KPZ equation (\ref{kpz}) in the stationary state.

The stochastic properties of the pair soliton growth mode is 
easily elucidated by noting that the action associated with the pair
mode is given by $S=(4/3)\nu\lambda|u|^3t$. Denoting the center
of mass of the pair mode by $x$ we have
$u=v/\lambda=x/t\lambda$ and we obtain using (\ref{dis}) the transition
probability
\begin{eqnarray}
P(x,t)\propto\exp\left(-\frac{4}{3}\frac{\nu}{\Delta\lambda^2}
\frac{x^3}{t^2}\right)~,
\label{rwpm}
\end{eqnarray}
for the `random walk' of independent pair solitons or steps in the
height profile. Comparing (\ref{rwpm}) with the distribution for
ordinary random walk originating from the Langevin equation
$dx/dt=\eta, \langle\eta\eta\rangle(t)=\Delta\delta(t)$,
$P(x,t)\propto\exp(-x^2/2\Delta t)$, we observe that the growth mode
performs anomalous diffusion. The distribution (\ref{rwpm}) also
implies the soliton mean square displacement, assuming pairs of
the same average size,
\begin{eqnarray}
\langle x^2\rangle(t)\propto
\left(\frac{\Delta\lambda^2}{\nu}\right)^{1/z}t^{2/z}~,
\label{msd}
\end{eqnarray}
with dynamic exponent $z=3/2$, identical to the dynamic exponent
defining the KPZ universality class. This result should be contrasted
with the mean square displacement 
$\langle x^2\rangle\propto\Delta t^{2/z}$,
$z=2$, for ordinary random walk. The growth modes thus perform
superdiffusion. 
\subsection{Scaling}
Aspects of the scaling properties of the noisy Burgers equation
are embodied in the scaling form (\ref{uscal}) for the slope
correlations. Here we give a set of heuristic arguments
implying that the dynamic scaling exponent $z$ can be inferred
from the exponent in the soliton dispersion law (\ref{dislaw}); we  refer to
\cite{Fogedby98b,Fogedby00a} for more details.

Within the weak noise approximation the slope correlations are 
given by appropriate overlap integrals involving the soliton
configuration \cite{Fogedby00a}. However, noting that the canonical
weak noise formulation in general follows from a saddle point
approximation to the Martin-Siggia-Rose functional integral,
we infer that the slope correlation can also be expressed as 
the time-ordered product \cite{Zinn-Justin89}
$
\langle u(xt)u(00)\rangle\propto
\langle 0|T\hat{u}(xt)\hat{u}(00)|0\rangle
$.
Here the ``quantum operators'' $\hat{u}$ and $\hat{p}$ evolve
according to the ``quantum Hamiltonian density'' (\ref{hamden})
and $|0\rangle$ denotes the zero-energy stationary state.
Displacing the field from $(x,t)$ to $(0,0)$, using the Hamiltonian and
momentum operators and inserting a complete set
of intermediate quasi-particle momentum states $|\Pi\rangle$,
we infer the spectral representation
\begin{eqnarray}
\langle u(xt)u(00)\rangle\propto\int d\Pi~G(\Pi)\exp{(Et-i\Pi x)}~.
\label{spec}
\end{eqnarray}
Here $G(\Pi)$ is an effective form factor and $E$ and $\Pi$ the
energy and momentum of the appropriate quasi particle.

The scaling limit for large $x$ and large $t$ corresponds to the
bottom of the quasi-particle spectrum and we note that only gapless
excitations contribute. Assuming a general dispersion law with 
exponent $\beta$, $E\propto\Pi^\beta$ the dynamic exponent $z$ is given
by the exponent $\beta$ for the
quasi-particle dispersion law. In the linear EW case the gapless
diffusive
dispersion law $E\propto \Pi^2$ yields the dynamic exponent
$z=2$; in the Burgers-KPZ case
the noise excites 
a new nonlinear gapless soliton mode with dispersion $E\propto
\Pi^{3/2}$,
yielding the exponent $z=3/2$. 
\section{Summary and conclusion}
We have here summarized recent work on the growth
morphology and scaling behavior of the noisy Burgers equation
in one dimension. Using a canonical weak noise approach to the
associated Fokker-Planck equation we have discussed  the growth
morphology in terms of a gas of nonlinear solitonic growth modes
with superposed linear modes. We have, moreover, associated
the dynamic scaling exponent with the soliton dispersion law.

So far the nonperturbative weak noise approach has only been 
implemented
in the one dimensional case where the analysis is tractable.
However, the weak noise method is generally applicable to 
generic Langevin equations driven by white noise  and it
remains to be seen whether the approach also throws light
on the higher dimensional case.

\end{document}